# Unsupervised Learning of Deep-Learned Features from Breast Cancer Images.

Sanghoon Lee
*Department of Computer Sciences and Electrical Engineering*
*Marshall University*
*Huntington, WV*
leesan@marshall.edu

Colton Farley
*Department of Computer Sciences and Electrical Engineering*
*Marshall University*
*Huntington, WV*
farley195@live.marshall.edu

Simon Shim
*Department of Biomedical Engineering*
*Marshall University*
*Huntington, WV*
shim@marshall.edu

Yanjun Zhao
*Department of Computer Science*
*Troy University*
*Troy, AL*
yjzhao@troy.edu

Wookjin Choi
*Department of Engineering and Computer Science*
*Virginia State University*
*Petersburg, VA*
wchoi@vsu.edu

Wook-Sung Yoo
*Department of Computer Sciences and Electrical Engineering*
*Marshall University*
*Huntington, WV*
yoow@marshall.edu

*Abstract*— Detecting cancer manually in whole slide images requires significant time and effort on the laborious process. Recent advances in whole slide image analysis have stimulated the growth and development of machine learning-based approaches that improve the efficiency and effectiveness in the diagnosis of cancer diseases. In this paper, we propose an unsupervised learning approach for detecting cancer in breast invasive carcinoma (BRCA) whole slide images. The proposed method is fully automated and does not require human involvement during the unsupervised learning procedure. We demonstrate the effectiveness of the proposed approach for cancer detection in BRCA and show how the machine can choose the most appropriate clusters during the unsupervised learning procedure. Moreover, we present a prototype application that enables users to select relevant groups mapping all regions related to the groups in whole slide images.

*Keywords—Breast cancer, whole slide images, machine learning, unsupervised learning, k-means clustering*

## I. Introduction

The advances in whole slide imaging technologies combined with the modern microscope equipment have generated a large volume of high-resolution digital images from microscopic slides providing highly reliable and useful image data in determining treatment outcome [1][2].

Several computerized approaches have been introduced to analyze whole slide images (WSIs) avoiding the tedious and error-prone manual process, and the digital image analysis approach based on machine learning techniques has shown successful results in detecting and diagnosing human cancers [3]. Machine learning-based approaches to analyze tissue samples obtained for the microscopic examination can provide satisfactory outcomes and validated clinical prediction, yielding software tools with onboard machine learning can assure faster turnaround times in a quantitative analysis of WSIs [4]. The machine learning-based approaches assist domain experts in providing definitive diagnoses and effective treatments in digital histopathology, promoting the transition of modern pathology.

Involving domain experts in validating machine learning models is common and is associated with the prediction of outcome. However, they still leave the potential human error or bias [5, 6]. Furthermore, labeling sufficient training samples to ensure accurate classification for hundreds-of-millions of histologic structures is very time-consuming and requires laborious effort, thereby causing a significant delay in the diagnostic process [7, 8]. The critical challenges in the WSI analysis using the machine learning-based approaches are 1) creating a methodology to reduce human efforts and 2) integrating machine learning capabilities into applications.

Unsupervised learning has been widely used in the field of machine learning, including clustering, density estimation, and pattern recognition. Unlike supervised learning where all training instances are required to be labeled, unsupervised learning does not involve training instances' labeling effort. The main advantage of the unsupervised learning approaches is that they are engaged to explore unknown datasets, and to discover label patterns and relationships in a given dataset, thus avoiding potential human error in the process of machine learning. However, unsupervised learning suffers from a few shortcomings: not very specific outcomes, existing outliers, and less accurate predictions.

In this paper, we propose an unsupervised learning method to detect cancer in breast invasive carcinoma (BRCA) images. BRCA images obtained from The Cancer Genome Atlas at the Genomic Data (TCGA) [9] are pre-processed through color normalization and square tessellation, transformed into deep-learned features by low-level feature characteristics. We used K-means clustering to assign the deep-learned features to K clusters as an unsupervised learning algorithm by measuring silhouette coefficients to determine the number of clusters for maximizing the qualifications of the neighboring clusters. The proposed approach is fully automated and does not require human involvement during the unsupervised learning procedure. Moreover, we present a prototype application for detecting cancer in whole slide images using the unsupervised learning approach. The contributions of our paper are that 1) a stratified unsupervised learning approach is proposed not only to avoid human effort but also to provide accurate prediction outcomes, and 2) a prototype cancer detection tool using the unsupervised learning approach is developed for whole slide image analysis.

The rest of the paper is organized as follows. In section 2, we explain well-rounded machine learning approaches to identify breast cancer in histopathological images and spatialized techniques adopted for breast cancer identification. In section 3, we describe the proposed unsupervised learning approach using deep-learned features extracted from whole slide images for breast cancer detection. In section 4, we present the preliminary results of the proposed method. Finally, conclusions and future works will be described in section 5.

## II. Related Works

Both supervised-learning and unsupervised learning have been successfully exploited in histopathological image analysis over a few decades [10]. Supervised learning methods aim to predict target sample images using labeled sample images, while unsupervised learning methods aim to find hidden patterns from unlabeled sample images. Artificial Neural Network (ANN) is a commonly used supervised learning method performing with three-layered neural networks: an input layer, a hidden layer, and an output layer. Based on the three-layered neural networks, unlabeled neurons in the output layer are determined. Even though ANN is suitable for non-linear multivariate problems, it has been mainly used for survival analysis [11, 12]. Another well-known supervised learning method is Support Vector Machine (SVM). SVM has been successfully adopted for breast cancer diagnosis because of its generalization capability avoiding an overfitting problem. However, SVM is infeasible for massive datasets and does not provide probabilistic explanations for the classification [13]. On the other hand, unsupervised learning methods have been used to find patterns on various types of datasets. K-means clustering is a well-known simple, yet powerful unsupervised learning method that partitions a collection of clusters into K cluster groups. K-means clustering scales to large datasets and generalizes well to unlabeled clusters, but the manual selection of K is an issue to overcome [14].

Meanwhile, computer-based image analysis has been more conveniently placed in the histopathological image interpretation. Specifically, representing the digitized images has led to substantial improvements and advances in image processing methodologies via pre-processing, segmentation, and feature extraction. Color normalization has been a critical stage in improving histopathology image analysis performance because of color variation in histopathology images [15]. Segmentation algorithms have been employed to obtain important prognostically related tissue specificity, and quantitative features have stimulated histological observations on biopsy specimens, solving the problem of observer variability and reducing the workload of pathologists. However, hand-crafted features designed before image segmentation require large computational tasks and consideration of tissue appearance (i.e., epithelial cancerous cell shows the highly irregular shape and enlarged size, remaining a considerable difficulty in identifying cancers).

Convolutional Neural Network (CNN) has received much attention with respect to its significant potential in the histopathological image interpretation [16]. While hand-crafted features rely on explicit algorithms requiring heavy efforts for novel datasets, features directly learned from raw data are trainable within CNN and can be automatically transformed in supervised-learning and unsupervised learning [17-19]. CNN features have provided better accuracy through convolution and pooling layers, followed by a fully connected layer for the learning task. In recent years, various innovative CNN models have been introduced in the field of machine learning communities. As a part of ImageNet project, VGG16 [20], a well-known CNN architecture consisting of 16 convolutional layers with lots of filters, has been widely used in image classification tasks [21]. In this paper, we utilize VGG16 to extract the deep-learned features from square tessellation regions.

## III. unsupervised Learning of Deep Learned Features from Breast Cancer Images

In this section, we describe the whole process of the proposed unsupervised learning of deep-learned features from breast cancer images. The overview of the whole slide imaging, segmentation, feature extraction, and clustering methods are shown in Fig 1.

### A. Whole slide imaging

A tissue section in histopathology image analysis is mounted on a glass slide through the tissue preparation. Then, it is stained based on tissue components (i.e., the nucleus is dyed with the blue or purple stain, while the cytoplasm is dyed with the pink stain by Hematoxylin & Eosin staining), and then converted to a digital image by a digital pathology slide scanner. Digitized histopathological images of tissues stained with different staining methods are finally prepared for computer-based image analysis. In this paper, we obtained seven whole slide images from TCGA. They are all hematoxylin and eosin-stained formalin-fixed paraffin-embedded (FFPE) sections of histopathology whole slide images of breast invasive carcinoma.

### B. Pre-processing

Color normalization was performed in the pre-processing step for whole slide images. First of all, we separated the foreground (tissue area) from the background using a two-component Gaussian mixture model to achieve the accurate differentiation between the tissue areas and the background of each whole slide image. The tissue area was partitioned into 2048 x 2048 x 3 sized tile images, and then normalized using Reinhard color normalization [15] . This normalization process uses the color distribution of the tissue areas mapped to that of the standard color images with the mean and standard deviation in LAB color space. The normalized tissue areas were segmented by square tessellation for 128 x 128 x 3 sized image patches. The pre-processing uses tools from HistomicsTK project and HistomicsML2 pipeline and its software [22, 23].

### C. Pipeline for feature extraction

128 x 128 x 3 sized image patches were resized to 224 x 224 x 3 sized image patches using nearest neighbor interpolation to represent input features for VGG16 [20]. VGG16 architecture comprises three types of layers: convolutional layers, max pooling layers, and fully connected layers. The convolutional layers consist of 13 layers with different number of filters. The number of filters in the first convolutional layer is 64, while the number of filters in the second convolutional layer is 128. VGG16 architecture was trained on the ImageNet dataset with 10+ million and 1000 classes. In this paper, we extracted the first fully connected layer representing a 4096-dimensional feature vector from VGG16 and performed an orthogonal linear transformation called Principal Component Analysis (PCA) to obtain 48 features [24]. We obtained 48 features from 4096 features to avoid latency and system overload during the process of the proposed approach.

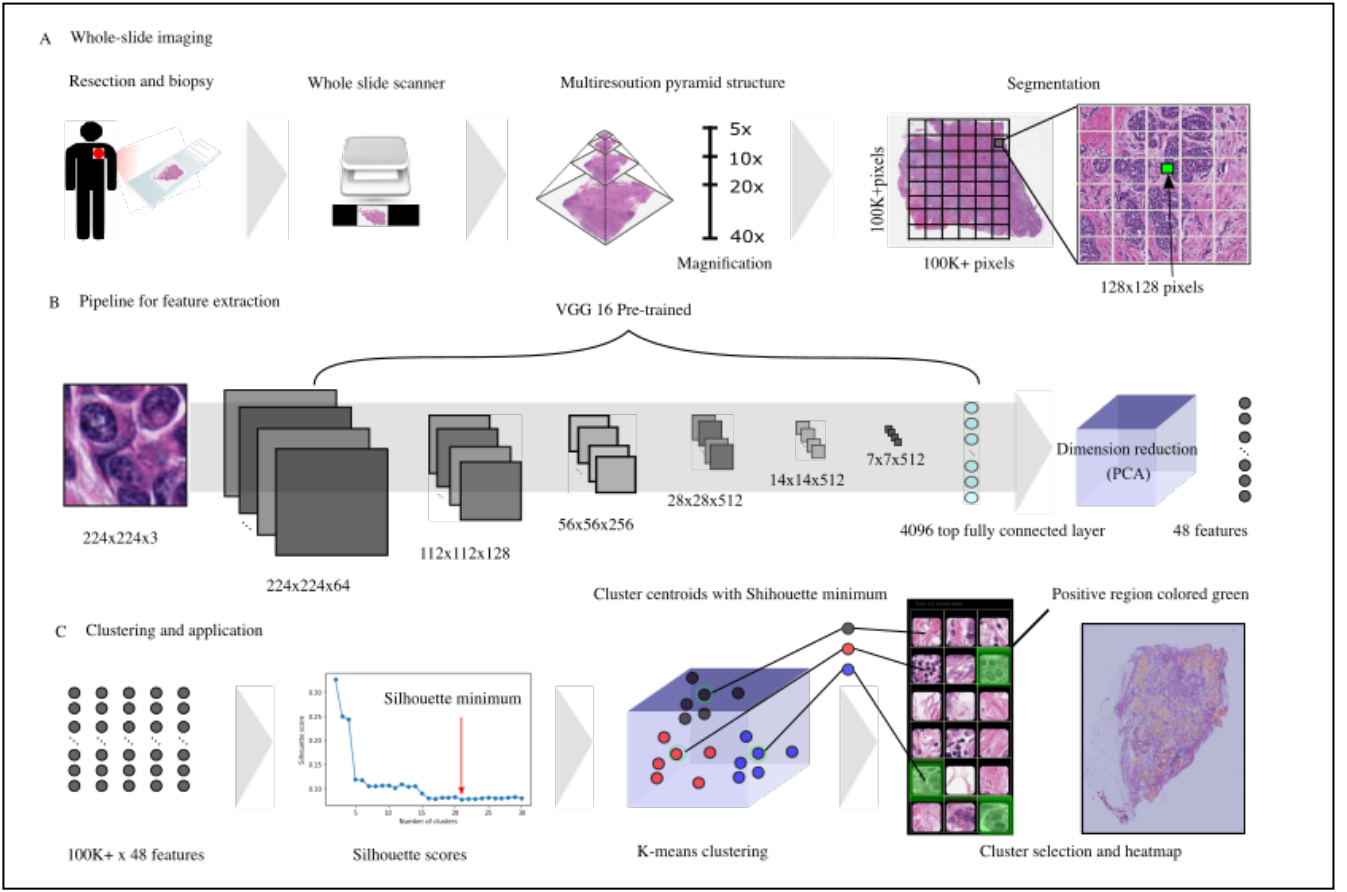

Fig. 1 Overview of the proposed approach and its application. Whole-slide imaging and segmentation steps are shown at the top. A pipeline for feature extraction through VGG16 architecture and a dimensional reduction technique are shown in the middle. A clustering procedure and its application are shown in the bottom.

### D. K-means clustering for K square tessellation region identification

A K-means clustering method to identify the K square tessellation region of the whole slide images is presented in this subsection. Forty-eight features extracted by the previous section represent a square tessellation region of whole slide images. In this paper, 100K+ square tessellation regions are finally extracted from the feature extraction pipeline described in the previous section. We used K-means clustering to partition the square tessellation regions into K distinct groups. However, determining the optimal number of clusters in the whole dataset is a typical problem for K-means clustering. In this paper, we computed the mean of silhouette coefficient [25] of all square tessellation regions to determine the optimal number of clusters. The average intra-cluster distance is computed as below:

$$N(x) = \frac{\sum_{y \in K_x, x \neq y} dist(x, y)}{|K_x| - 1} \quad (1)$$

For each square tessellation region $x \in K_x$ in the cluster $K_x$, the mean of intra-cluster distance $N(x)$ is computed as the sum of $dist(x, y)$, the distance between $x$ and $y$ tessellation region excepting the case where $x$ and $y$ are same, divided by $|K_x| - 1$. The mean of the nearest cluster distance is computed as below:

$$M(x) = \min_{z \neq x} \frac{\sum_{y \in K_z} dist(x, y)}{|K_z|} \quad (2)$$

For each square tessellation region $x \in K_x$ in the cluster $K_x$, the mean of the nearest cluster distance $M(x)$ is computed as the smallest value represented as the sum of $dist(x, y)$, the distance between $x$ and $y$ square tessellation region where $x$ is not same with $z$, divided by $|K_z|$. Thus, a silhouette coefficient of a square tessellation region is defined as below:

$$\text{silhouette}(x) = \frac{M(x) - N(x)}{\max(M(x), N(x))} \quad (3)$$

, where $|K_x| \geq 2$.

K-means clustering as a centroid-based algorithm is performed by calculating the distance between two square tessellation regions after obtaining the optimal number of clusters in the silhouette coefficient equation. K-means clustering is a simple unsupervised learning algorithm that partition a dataset into K pre-defined clusters. The algorithm is fast, relatively efficient, and even very easy to understand. The assumption is that the dataset is well-separated from each other. In this paper, we use the K-means clustering algorithm to partition square tessellation regions into K groups. In addition to K-means clustering, we identify K representative square tessellation regions which are the closest to the centroid of the groups. In this paper, we use a typical squared error function as an objective function given by:

$$J(C) = \sum_{i=1}^{k} \sum_{j=1}^{k_i} (||s_i - c_j||)^2 \quad (4)$$

, where $k$ is the number of clusters and $k_i$ is the number of regions that belong to the $i$th cluster. Besides the clustering objective function, K-means clustering uses an equation for identifying each cluster center over the iterative process. A cluster center $c_i$ is computed as below:

$$c_i = \frac{\sum_{j=1}^{k_i} s_i}{k_i} \quad (5)$$

K means clustering algorithm to find K representative square tessellation regions follows:

---

**Algorithm 1** K means clustering algorithm to identify the representative K square tessellation regions

---

**Input:** S = $\{s_1, s_2, s_3, ..., s_n\}$ be the set of square tessellation regions and C = $\{c_1, c_2, c_3, ..., c_k\}$ be the set of centers

**Output:** K representative square tessellation regions

1: Select k cluster centers randomly.
2: For every region in S, compute the distance between the region and cluster centers using (4).
3: Find the minimum distance and assign the region to the cluster center.
4: Compute cluster centers using (5).
5: Repeat from Step 2 until no regions found to assign.
6: For every region that belongs to each cluster in S, compute the distance between the region and the cluster centroid.
7: Find the minimum distance and assign the region to the representative region for the cluster.

---

The output of Algorithm 1 for identifying K representative square tessellation regions is shown in Fig. 2. 'TCGA-OL-A66I-01Z-00-DX1' represents the name of the whole slide image, and the group names are represented by the numbers, 7, 14, and 12. Each group includes a representative square tessellation region directly linked to other square tessellation regions with the same cluster. The square box colored as blue at the top of the figure indicates the representative square tessellation region of group 7. The square box colored as red in the middle shows the representative square tessellation region of group 14, while the square box colored as blue at the bottom of the figure indicates the representative square tessellation region of group 12.

K-means clustering algorithm to the K representative square tessellation regions identification provides insights into an implementation of a prototype web-based application that enables domain experts to determine the K representative square tessellation regions so that clustered areas can be detected without further consideration. The web application is built upon HistomicsML2 to use the interactive learning process avoiding selection errors of the K-means clustering algorithm. Users can choose the most appropriate samples with the closest centroids applied to the entire dataset. The interactive process is performed over several steps and classifies all square tessellation regions of WSIs as positive or negative. The web application includes a heatmap function for the positive class density determined by the K-means clustering algorithm.

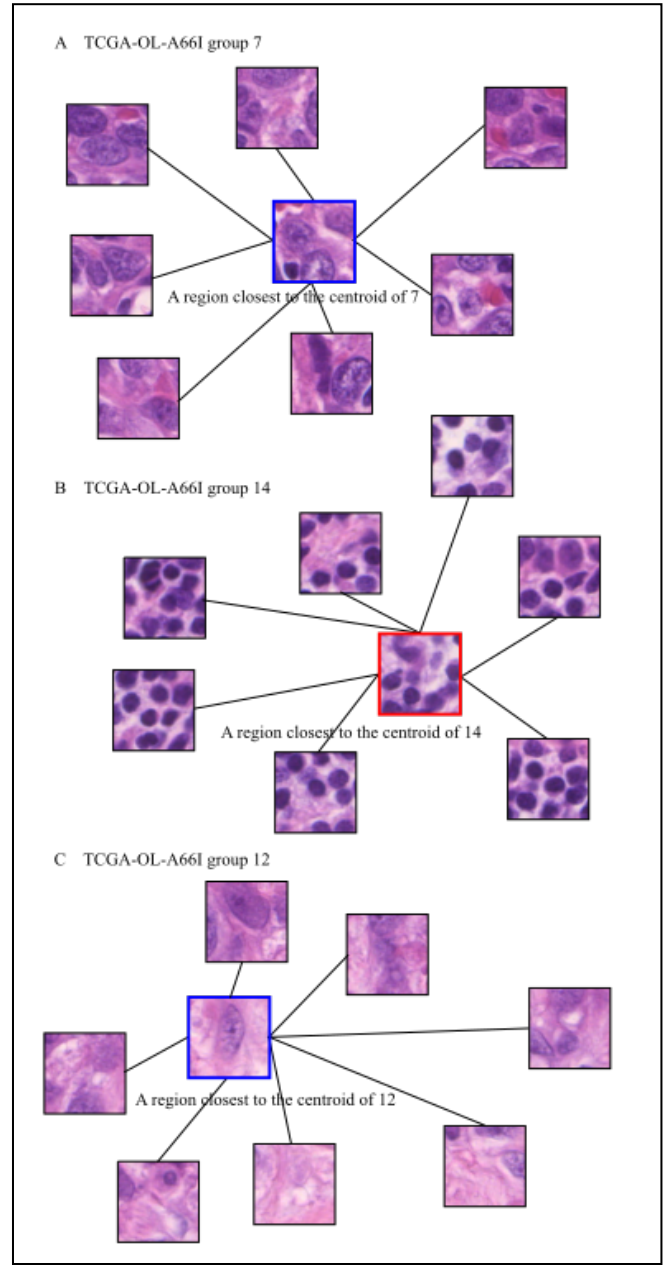

Fig. 2 Three representative square tessellation regions that belong to each group 7, 14, and 12 respectively. Square tessellation regions linked to the regions colored as blue in the top and the bottom belong to the groups of 7 and 12 respectively, while other square tessellation regions linked to the region colored as red in the middle belong to the groups of 14.

The web application follows a few-shot learning strategy, a type of machine learning methodology that limits the number of training samples in a supervised manner but generalizes to new samples well. This machine learning methodology can be categorized by augmenting training samples, constraining hypothesis space, or altering search strategy in hypothesis space [26], helping to relieve the burden of human efforts in a large amount of data. Because WSI analysis requires domain experts to provide sufficient annotated samples and very laborious work, it is worth considering the few-shot learning strategy in the whole slide image analysis. Furthermore, few-shot learning can minimize the data collection effort for users to organize their classifiers so that the entire process of analyzing whole slide images via applications can be done as quickly as possible. Thus, the interactive process includes a multi-sample selection to determine whether the square tessellation regions are positive or negative.

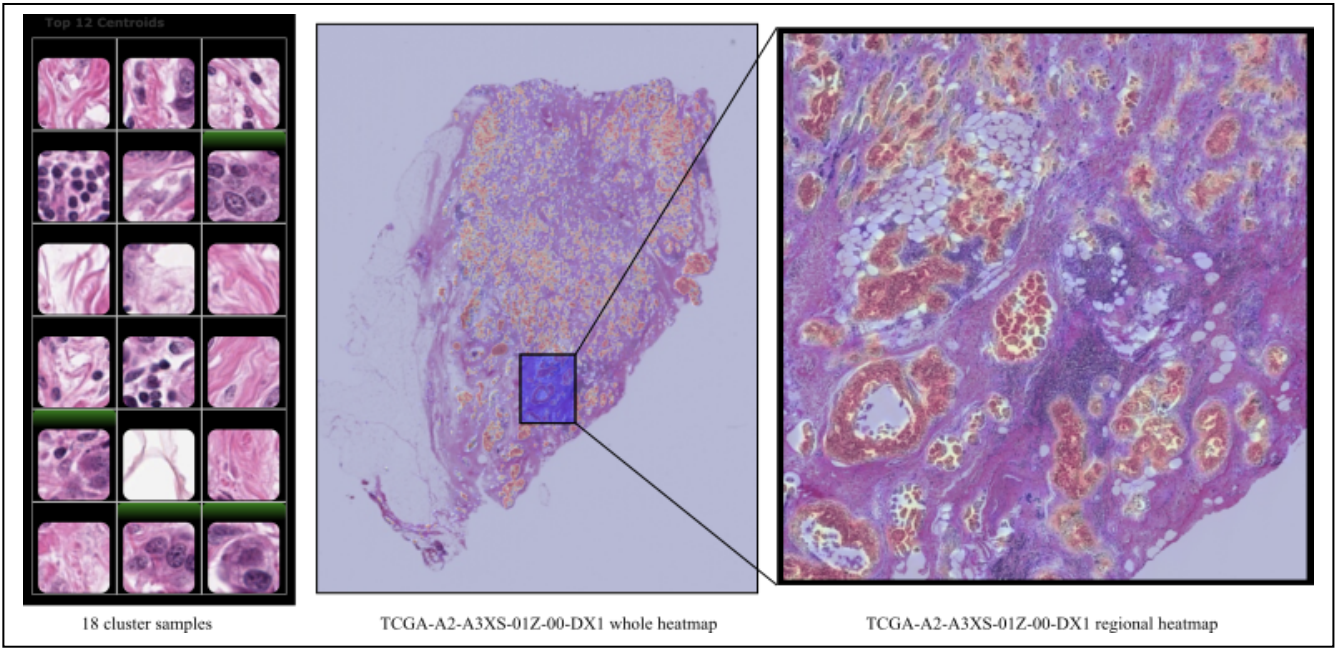

Fig. 3 The prediction resutls on the web application using a few-shot leaning strategy. Eighteen square tessellation regions closest to the centrods of the clusters are shown in the left. A heatmap generated from the web application is shown in the middle. A higher zoom level view is shown in the right.

An example of our application following a few-shot learning strategy is shown in Fig 3. 18 selected square tessellation regions nearest to each cluster centroids are shown in the left figure. The labels colored as green represent positive cancer, and others represent negative cancer. A heatmap for a whole slide image is shown in the middle, and more details are shown in the right. The prototype of our web application is open source and prepared as a docker image.

## IV. EVALUATION

In this section, we present the experimental results of the proposed approach for detecting breast cancer in whole slide images. We describe the dataset, evaluation metrics, and prediction results used for the experiments in the section.

### A. Dataset

Seven Breast invasive carcinoma (BRCA) whole slide images obtained from The Cancer Genome Atlas at the Genomic Data (TCGA) [9] were pre-processed, segmented, and generated 1,627,434 square tessellation regions. Regions of interests (ROIs) were annotated for each BRCA whole slide image by two experts.

### B. Evaluation Metrics

The evaluation for detecting the cancer regions in the BRCA images was performed based on the accuracy and F1-score. We use the accuracy and F1-score metrics as below:

$$Accuracy = \frac{(TP + TN)}{(TP + TN + FP + FN)} \quad (6)$$

$$F_1 = \frac{2 \cdot TP}{2 \cdot TP + FP + FN} \quad (7)$$

, where TP represents the true positive, the number of positive predictions actually positive. TN represents the true negative, the number of negative predictions correctly identified as negative. FP represents the false positive, the number of negative predictions incorrectly identified as positive. FN represents the false negative, the number of positive predictions incorrectly identified as negative. The results of the evaluation are shown in Table 1. Each row represents the results, including Silhouette optimal number, cluster groups, accuracy, and F1-score for each BRCA whole slide image. The Silhouette optimal number was obtained by following (1), (2), and (3). These numbers were used as the number of clusters to perform K means clustering. The cluster set is the list of clusters used for measuring the accuracy and F1-score. For example, 26 clusters obtained by minimizing the Silhouette scores were set for the initial number of clusters when performing K-means clustering the BRCA whole slide image named 'TCGA-E2-A1LS-01Z-00-DX1'. We finally selected two representative square tessellation regions belonging to the 18th and the 5th clusters to measure the accuracy and F1-score. The average accuracy and the average f1-score on the dataset are 0.8454 and 0.8293 respectively

### C. Prediction results

The representative prediction results of whole slide images (TCGA-OL-A66I-01Z-00-DX1, TCGA-A2-A0YM-01Z-00-DX1, TCGA-A2-A3XT-01Z-00-DX1, and TCGA-A7-A0DA-01Z-00-DX1) are shown in Fig 4. We selected 4 representative prediction results among 7 slides because they provide better performance than others. Some extensive research works are required to resolve the issue with low performance. The first column represents the original image resized from 40x magnification to 5x magnification. The predicted regions of cancer colored as blue and tumor-infiltrating lymphocytes colored as red are in the second column. The third column represents the cancer heatmap predicted by the proposed method. The tumor-infiltrating lymphocytes' heatmaps predicted by the proposed method are in the last column. The grid regions of each heatmap are sized as 40 x 40. The results shown in Fig 4 indicate that TILs regions are more likely to be located around the corner from the whole slide images surrounding the cancer regions. Analyzing spatial relationships between cancer and the tumor-infiltrating lymphocytes benefits the cancer research community, but we leave it to future research to demonstrate the effectiveness of the proposed approach.

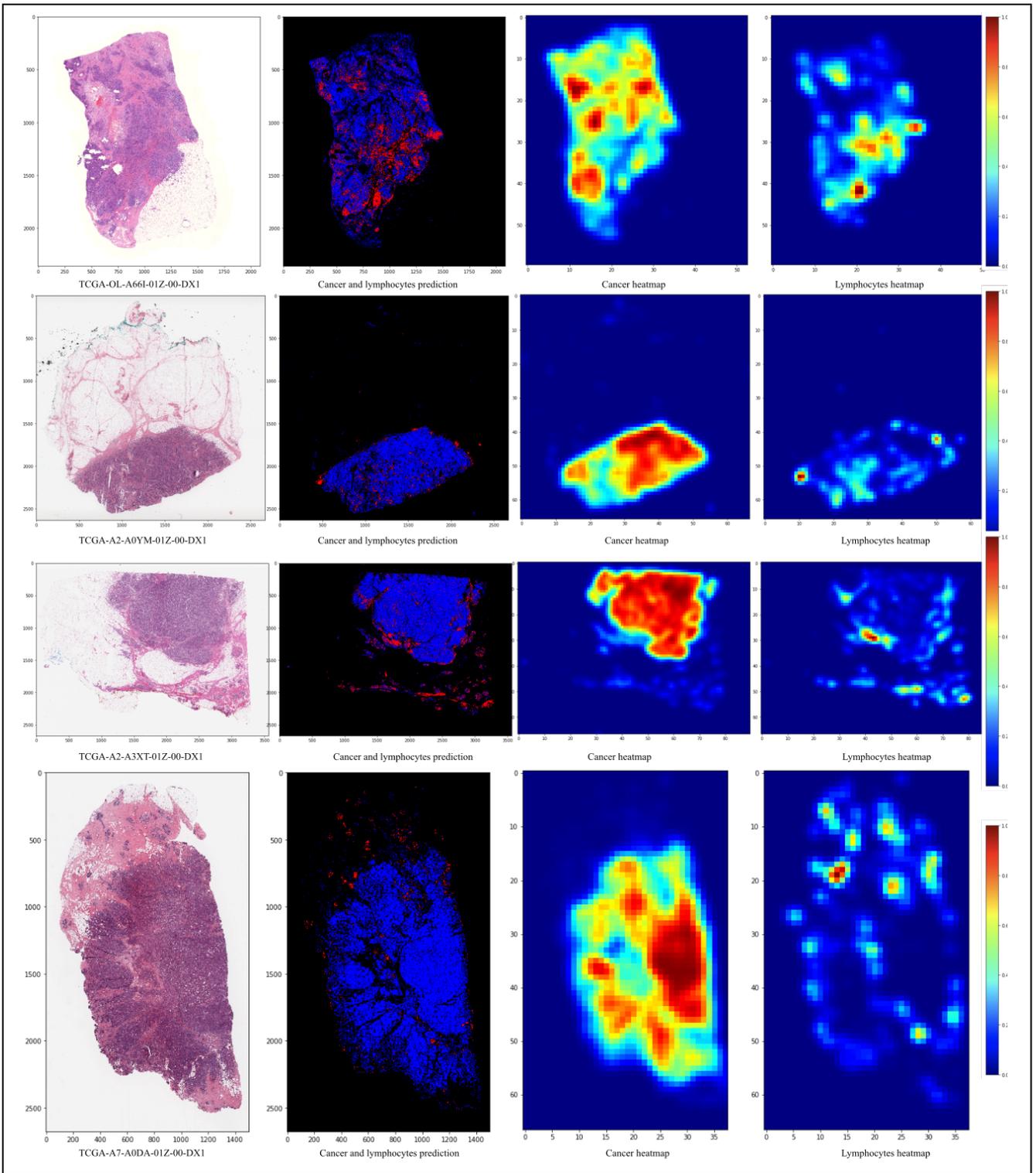

Fig. 4 Prediction results for four whole slide images on 40x magnification. The first column represents the original whole slide image, the second column represents the prediction results (cancer colored as blue while tumor infilterating lymphocytes colored as red), the third column represents a heatmap for cancer regions, and the last column represents a heatmap for tumor infiltrating lymhocytes regions. The whole slide images at the first two columns are resized to 5x magnfication, while the heatmap images at the last two columns are generated in 40 x 40 grids.

## V. CONCLUSIONS AND FUGURE WORKS

In this paper, we proposed an unsupervised learning approach for breast cancer detection in whole slide images. Whole slide images were segmented into square tessellation regions after pre-processing, and deep-learned features were extracted from the square tessellation regions through a deep learning model architecture with a dimensionality reduction technique. Silhouette coefficients were measured to determine the number of clusters, and the K-means clustering algorithm is adopted to obtain K square tessellation regions in the whole slide images. In addition to the description of the proposed approach, the performance evaluation was conducted in two well-known metrics: accuracy and F1-score. The averages of the accuracy and F1-score are 0.8454 and 0.8293 respectively over the given dataset.

TABLE I. CANCER DETECTION ACCURACY AND F1-SCORE

| Slide name | Silhouette optimal number | Cluster set | Accuracy | F1-score |
|---|---|---|---|---|
| TCGA-A7-A0DA | 29 | [25, 22, 6, 2, 24, 14, 0, 20, 10] | 0.8829 | 0.8929 |
| TCGA-A2-A0YM | 20 | [7, 6, 1, 5] | 0.8360 | 0.8863 |
| TCGA-A2-A3XT | 19 | [13, 10, 5, 1, 2] | 0.9316 | 0.9514 |
| TCGA-BH-A0BG | 8 | [1, 5] | 0.7828 | 0.6857 |
| TCGA-E2-A1LS | 26 | [18, 5] | 0.8495 | 0.8680 |
| TCGA-OL-A66I | 25 | [7, 18, 12, 20, 1, 0] | 0.7761 | 0.7091 |
| TCGA-C8-A26Y | 21 | [9, 18, 12, 16] | 0.8594 | 0.8122 |

Moreover, we implemented a prototype web-based application enabling users to choose appropriate square tessellation regions without the laborious efforts. The proposed approach is fully automated and does not require further inputs during the unsupervised learning procedure. A heatmap is displayed with the corresponding clusters when the user selects the square tessellation regions closest to the centroids of the clusters.

Recent studies have shown that statistical approaches for analyzing spatial characteristics in histopathological images play an important role in the tumor microenvironment. These studies suggest a new approach to the issues in spatial analysis in whole slide images. We plan to investigate and compare different spatial analysis methods in the future.